 \documentclass[12pt,preprint]{aastex}

\usepackage{amsmath}
 \usepackage{color}

\shorttitle{The evolution of emission lines}

\shortauthors{Liu \& Mao}

\begin{document}


\title{Non-detection of the Gamma-ray Burst X-ray Emission Line: The Down-Comptonization Effect}


\author{Jie-Ying Liu\altaffilmark{1,2,3} and Jirong Mao\altaffilmark{1,2,3}}

\altaffiltext{1}
{Yunnan Observatories, Chinese Academy of Sciences, Kunming 650011, Yunnan Province, China}
\altaffiltext{2}
{Center for Astronomical Mega-Science, Chinese Academy of Sciences, 20A Datun Road, Chaoyang District, Beijing 100012, China}
\altaffiltext{3}
{Key Laboratory for the Structure and Evolution of Celestial Objects, Chinese Academy of Sciences, Kunming 650011, China}

\email{jirongmao@mail.ynao.ac.cn}

\begin{abstract}
The detection of the gamma-ray burst (GRB) X-ray emission line is important for
studying the GRB physics and constraining the GRB redshift. Since the
line-like feature in the GRB X-ray spectrum was first reported in 1999,
several works on line searching have been published over the past two decades.
Even though some observations on the X-ray line-like feature were performed,
the significance remains controversial to date.
In this paper, we utilize the down-Comptonization mechanism and present the time evolution of the Fe K$\alpha$ line emitted
near the GRB central engine.
The line intensity decreases with the evolution time, and the time evolution depends on the
the electron density and the electron temperature.
In addition,
the initial line with a larger broadening decreases less over time.
For instance, when the emission line penetrates material with the an electron density above
$10^{12}$\,cm$^{-3}$ at 1 keV, it generally becomes insignificant enough after $100$\,s for it not to be detected. The line-like profile deviates from the Gaussian form, and it finally changes to be similar to a blackbody shape at the time of the thermal equilibrium between the line photons and the surrounding material.
\end{abstract}


{Unified Astronomy Thesaurus concepts:} Gamma-ray bursts(629); X-ray detectors(1815);

\section{Introduction}
Comptonization is the energy transfer process by the scattering between the photons and plasma.
This changes the strength and shape of the radiation spectrum, and also causes changes to the
  plasma temperature. If the average photon energy is less than the electron thermal
  energy, the photon frequency will move to a higher energy during
  the scattering. This radiation process is adopted to study the gamma-ray burst (GRB) emission
   \citep[e.g.,][]{Ghisellini+1999,Meszaros+2000}.
 In contrast, if the average photon energy
 is larger than the electron thermal energy in a cold environment, down-Comptonization will take place.
  The evolution behaviors of
 the different incident spectra, including the continuum and the emission line
  under the down-Comptonization effect, was studied \citep{Chen+1994,Deng+1998,Liu+2004}.
  The down-Comptonization process has also been used to
  interpret the broadening and redshift of the line emission
  \citep{Ross+1978,Czerny+1991,Misra+1998}.

\citet{Kallman+2003} performed high-energy $\gamma$-ray softening to hard X-rays. When
 the incident hard X-rays illuminate a cold ion-rich environment
 at about $10^6-10^7$\,K (also called a reprocessor), the emission line will
 be generated through fluorescence or a recombination mechanism \citep[e.g.,][]{Ross+1978, Fabian+2000, Reeves+2001,Reynolds+2003}.
They used this mechanism to interpret the GRB line-detection.  According
  to the mechanism for the iron-line production, the density, composition,
 location, and bulk motion of the reprocessor are constrained by
 the line-like feature detection. Therefore, the detection of the GRB emission
 line in the X-ray band is important for studying
GRB physics, and it is also helpful for GRB redshift determination.

Since the iron-line was first observed in GRB 970508 by BeppoSAX \citep{Piro+1999},
 some works on line-like feature searching were published. Before the launch of Swift,
 the line-like features were identified in a few GRBs \citep{Piro+1999,Yoshida+1999,Antonelli+2000,Piro+2000,Yoshida+2001,
  Watson+2002,Watson+2003,Butler+2003,Bottcher2003,Bottcher2004}.
Most of the line features were recognized as emissions, and they were observed about several hours to
  1 day after the GRB trigger. The absorption
lines were only observed in the early observing times ($t<20$\,s for GRB 990705 in Amati et al. (2000), and $t<300$\,s
for GRB 011211 in Frontera et al. 2004). Excepting a significance of 4.7\,$\sigma$
in GRB 991216, other line features are almost all less than 3.0$\sigma$. \citet{Gou+2005}
compared the probability of the iron-line detection in GRB X-ray afterglow using Swift-XRT, Chandra and
 XMM-Newton. Due to the differences in the effective area and the slew
  response time of these instruments, they predicted that Chandra and
  XMM-Newton would detect the iron-line emission with a larger
   significance at more depth and a later time than Swift. They deduced
    that Swift can distinguish a high-significance level up to 5$
\sigma$ between a broad and narrow line out to $z\lesssim5.0$ for times
of $t\lesssim 10^4$\,s.  However, after the launch of Swift in 2004,
there was little optimism around line searching. Only two line-like features
were reported. \citet{Butler+2007} suggested that the line emission in
GRB 060218 is due to L-shell transitions of Fe at roughly solar abundance.
The possible line emission occurs near $t\sim 1$\,ks and the emitting material
is at $R\sim10^{13}$\,cm. The Gaussian component was added to fit GRB 060904B,
which is considered as the line emission of the highly-ionized nickel
recombination \citep{Margutti+2008,Moretti+2008}.

 Not only has the number of detections of the line-like features not dramatically increased, more and more works doubt the significance of their detection.
  \citet{Borozodin+2003} claimed that no evidence
    for spectral lines were present in GRB 011211. \citet{Watson+2003}
    showed the highest level of the confidence for the X-ray line detection (Si XIV with
     $3.7\sigma$) in GRB 030227 but nondetection of Fe, Co,
      and Ni emission lines. \citet{Sako+2005} performed detailed X-ray
      spectral analysis of 21 bright GRB afterglows including
       nine sources in which the line-like features were reported in previous works. They
        had some doubts on the significance reported before. \citet{Hurkett+2008}
        thoroughly compared three methods applied in the literature
        claiming the significance of the line-like features. They
        failed to find evidence for strong lines through analysis of 40
        GRBs with enough photon counts in the Swift-XRT windowed timing mode.
        \citet{Giuliani+2014} did not find any significant lines in GRB 120711A. \citet{Campana+2016}
          focused on the searching for the narrow emission/absorption features in the X-ray
          spectra observed by XMM-Newton.
           They concluded that the line features are not related to the GRB
            environment but are most likely from a Galactic origin.

In general, even though some works have claimed detection
of the line-like features, their significance is still under debate.
The physical issues that make the emission line only marginally detected or even not observed are
focused on in this paper. We utilize the down-Comptonization mechanism to explain the nondetection of
the GRB emission line in the X-ray band. The original emission line can interact with
the low-temperature electrons, and the down-Comptonization process decreases the line strength, and deviates the line profile from the Gaussian form.
The emission line finally changes to be a shape similar to the blackbody at the time of the thermal
  equilibrium between the photons and the environment.

The down-Comptonization process was mathematically described by the extended Kompaneets equation \citep{Chen+1994}.
   We show the model in detail in Section 2. The effects from
the density and temperature of the environment
on the line evolution are shown in Section 3. In addition, the evolution of the initial emission lines
with two different broadenings are also compared.
 We discuss our results and the implications
in section 4. The conclusion is presented in Section 5.

\section{Model}\label{model}
 When the surrounding environment is not very dense, the photons may undergo only little scattering. For each scatter, the change of the photon energy is, at most, on the Compton wavelength, or $\sim70$\,eV at the energy of the iron-line. Therefore, the scattering is discrete, and the shape of Compton scattering kernel is apparent in the scattered-line profile \citep{Pozdnyakov+1979}. However, if the line photons pass through the very-dense region, then significant scattering will dramatically affect the line profile, causing it to deviate from the Gaussian profile under the Doppler broadening and recoil effect. Even though the change of photon energy in each scattering is discrete, overall the large number of photons under multiple scattering will make the energy change continuous. Kompaneets described the interaction between nonrelativistic matter
 ($kT_e\ll m_ec^2$) and low-energy radiation ($h\nu \ll m_ec^2$) as the mixture
 of photons and electrons \citep{Kompaneets+1957}. The change of the radiation
 field caused by the scattering between the photons and electrons is considered
 as the diffusion of the photons in the frequency space.
 In this paper, we concentrate on the evolution of the emission line
 in the cool ambient material around line-production region. The physics of the GRB environment is very complicated. In particular, along with the jet moving forward, the distributions of density and chemical composition change in the GRB jet. To simplify this, we suppose that the environment of the line-emitting region is isotropic and the density is larger than the typical value of wind.
 Given a certain fireball radius $R$, the medium that is swept by the forward shock is fixed in a certain region with a thickness of $R/\Gamma^2$, where $\Gamma$ is the bulk Lorentz factor. Thus, the energy change described by the Kompaneets equation is in a certain volume.
 The spatial transport of radiation is important when we have only a few scattering times. However, we assume that the scattering is continuous in this paper, and the spatial transport of radiation can be neglected.
 This down-Comptonization process
 can be described by the extended Kompaneets equation \citep{Chen+1994},
 which reads
\begin{equation}\label{e_kompaneets}
\frac{\partial n}{\partial t}=\frac{kT_{e}}{m_{e}c^2}n_{e}\sigma_{T}c\frac{1}{x^2}\frac{\partial }{\partial x}\left\lbrace x^4\left(1+\frac{7}{10}\frac{ kT_e}{m_{e}c^2}x^2\right)\left(\frac{\partial n}{\partial x}+n^2+n\right)\right\rbrace
\end{equation}
where $n_{e}$ and $T_e$ are the number density
 and the electron temperature, respectively, and
$x\equiv\frac{h\nu}{k T_{e}}$ is the dimensionless photon frequency.
 $n(x,t)\equiv n(\nu,t)$ is the photon occupation number of the radiation field, which represents the photon number in each volume at a
 frequency of $\nu$. The former extended Kompannets equation deduced by \citet{Ross+1978} can describe the energy change well. Although the equation is slightly different to Equation (1), they are consistent for the photon energy that is considered in this paper.
For example, Liu et al. (2004) adopted Equation (1) and obtained the result, which is similar to that from \citet{Ross+1978}.

The term $\left(\frac{\partial n}{\partial x}+n^2+n\right)$ on
 the right-hand side of Equation (1) guarantees the thermal distribution as $n=(e^x-1)^{-1}$ when
  photons and electrons reach the thermal
 equilibrium. Equation (1) can be applied to wide photon frequencies, i.e.,
  no matter whether the averaged photon energy $h\bar{\nu }$ is larger than $kT_{e}$ or not.

 The profile of the original incident emission line
 is set to be Gaussian, and the photon occupation number can be written as
 \begin{equation}
n(x,0)= x^{-3} {\rm exp} \left[-\frac{\rm{4 ln2}}{(\Delta x)^2}(x-x_0)^2\right].
\end{equation}
 The central energy of the emission line corresponds to the energy
 of the Fe K$\alpha$ in the rest frame, i.e., $x_0=h\nu_0/kT_e=6.4\,{\rm keV}/kT_e$.
 The FWHM $h\Delta\nu$ is equal to $\Delta x\cdot kT_e $.

  According to the distance of the reprocessor
  to the central GRB, the theoretical line-production models can be subdivided into the engine model
   and the geometry model \citep[e.g.,][]{Lazzati+1999,Vietri+2001,Reeves+2002}. These two scenarios suggest that the reprocessors are located in $10^{13}-10^{16}$\,cm.
 The density in the radial distance ranges as $10^{11}-10^{17}$\,$\rm cm^{-3}$ \citep{Kallman+2003}.
  In this paper, the density of the environment
   interacting with the original emission line is assumed to be
    uniform and less than $10^{17}$\,$\rm cm^{-3}$. We also set a lower limit that is
    $10^{6}$\,cm$^{-3}$ as a typical value of the GRB wind environment \citep{Chevalier+1999}.
    During the calculation, the temperature $kT_{e}$ is set to be 1\,keV.

     FWHM is one of  the indicators of line broadening.
     Supposing that the original emission line
 is also broadened by the bulk motion in the reprocessor, we can obtain the
\begin{equation}
\Delta{\nu}\sim\frac{\nu_0}{c}\sqrt{V_{\rm bulk}^2}=\frac{\nu_0}{c}\times V_{\rm bulk}=\beta\nu_0,
\end{equation}
  where $\beta=V_{\rm bulk}/c$. Some research on the line detection
  suggests that the X-ray plasma flow comes out
   at a velocity $\sim 0.1$c \citep{Reeves+2002,Butler+2003}.
   Therefore, we set the characteristic FWHM as $ h\Delta\nu=0.1h\nu_0$.

 \section{Results}\label{results}

We use the fully implicit difference scheme to solve the equation
 (\ref{e_kompaneets}). In the top-left panel of Figure 1, we present the evolution of
 the emission line with $\Delta \nu/\nu_0 =0.1$ in the case of
 $kT_e=1.0$\,keV and $n_e=2.0\times10^{8}$ cm$^{-3}$.
 The line strength decreases with the time
 because the energy is transferred to electrons. 
 At $t\sim 10^4$\,s, the line intensity is about 95\% of the initial intensity.
 At $\sim 10^5$\,s, it decreases to 70\%. After
 $t=10^7$\,s, when the thermal equilibrium between the plasmas and
 emission photons is approaching, the spectral profile looks like a blackbody
 continuum with $kT_B\sim1.0$\,keV. Moreover, the intensity peak moves toward to a lower energy,
 which is also called as the Compton-redshift in some literature.

 In order to compare the electron density that takes effect on the line intensity, we show the line evolution in the case of $n_e=2.0\times10^{12}$\, $\rm cm^{-3}$ in the top-right panel of Figure 1. We note that the evolved-line profile is similar to that in the case of $n_e=2.0\times10^{8}$\, $\rm cm^{-3}$ at different timescales. This is because the evolution of the incident spectrum is determined by the electron number density and the electron temperature in the Comptonization process. For the same
 temperature, the scattering timescale is proportional to $1/n_e$, as shown in Figure 1. This denser environment increases the probability of scattering between the photons and electrons. Thus, the line intensity decreases more quickly. The peak intensity decreases to 30\% at about 100\,s in the case of $n_e=2.0\times10^{12}$\,cm$^{-3}$.

 Besides the decreasing intensity, we note another dramatic evolution result, i.e., the
emission-line profile deviating from Gaussian form with time.
The strong deviation begins at $t\sim10^7$\,s for $ n_e=2\times10^{8}$\,$\rm cm^{-3}$, while for $ n_e=2\times10^{12}$\,$\rm cm^{-3}$, this deviation will begin as early as $10^2$\,s.
It is shown in our results that the emission line is hard to detect after $10^2$\,s.
     Although Swift-XRT can perform observation during the prompt-emission phase
      when we have an alert from a GRB precursor, a normal observation in the X-ray
      band begins 50-100\,s after a GRB trigger. This indicates that the effective
       detection to the emission line should be performed earlier than 100\,s in
       case the electron density be less than about $10^{12}$\,${\rm cm^{-3}}$.
        The nondetection of the emission line by Swift-XRT implies that the
        electron density in the reprocessor is very dense. In addition, the signal-to-noise
        of the emission line has a contrast to the continuum. When the GRB X-ray flux
        decreases very quickly in total, X-ray emission lines are sustained in a shorter time.
        Thus, it is even harder to detect X-ray emission lines.

      Setting the same density of $n_e=10^8$\,$\rm cm^{-3}$, we show the temperature effect
      on the evolution in Figure 1. In the bottom-left panel, the
temperature is 3\,keV and the line evolution is similar to the case of $kT_e=1.0$\,keV.
Corresponding to the same time, the line intensity decreases by more than that
 at a lower temperature. The redshift of the line center is less because
 of less temperature discrepancy between photons and the surrounding medium.

 The effect of FWHM of the line with $\Delta\nu/\nu_0 =0.5$ is shown in the bottom-right panel of Figure 1. The photon occupation number is
 \begin{equation}
n_2(x,0)= A \cdot x^{-3} {\rm exp} \left[-\frac{\rm{4 ln2}}{(0.5x_0)^2}(x-x_0)^2\right].
\end{equation}
Integrating the intensity of this emission line and making sure it is the same as that in equation (2), we obtain the value of $A$ as 0.2. Thus, the peak intensity of the initial line is normalized to 0.2 at the center of the emission line. In this case, the line center gradually moves to the lower energy with time, but the peak intensity only decreases a little. Even at $t=10^{6}$\,s, the peak intensity does not change much. The reason is that the time evolution is mainly determined by the term of $\frac{\partial n}{\partial x}$ when $n(x,0)$ is small. The value of $\frac{\partial n}{\partial x}$ is larger for the narrower line cases.
This indicates that the wider emission lines can be easier identified than the narrow lines. However, because of the same total initial-line intensity, and by passing through the same scattering depth, the change of the intensity in each emission line is the same in the same timescale. The line photons will finally have an equilibrium with the surrounding matter through sufficient scattering. The results can be clearly seen in Figure 1.

  \section{Discussion}
We theoretically investigate the down-Comptonization effect on the GRB emission line in the X-ray band. From the observational point of view,
current detectors have limited sensitivity. Therefore, it is hard to detect some emission lines in GRB X-ray afterglows.
On the other hand, we expect the emission lines produced in the environment with strong metal abundances, which are much greater than solar ones. A large number of X-ray photons are also required in GRB early afterglows to generate strong emission lines. Some future satellites, such as SVOM, EP, THESEUS, and HiZ-GUNDAM, are planned to effectively perform the detection.

In this work, we obtain the evolution of the emission line penetrating the
  medium surrounding the line-production regions. According to the density range in
  the radial distribution in \citet{Kallman+2003}, we compare the intensity and the profile of the emission line in two cases of the different electron number densities less than $10^{17}$\, cm$^{-3}$. We take $n_e=10^8$\, cm$^{-3}$ and $n_e=10^{12}$\, cm$^{-3}$, for instance. The case with the electron number density
    of $n_e=10^6$\,$\rm cm^{-3}$
    can exist as the stellar-wind case \citep{Chevalier+1999}.
       The cases with the higher number density indicate that the surrounding materials exist
     closer to the central engine, or that the materials are inside the GRB jets
     launched by the central engine.
     We normally prefer the reprocessor materials with a reasonable electron
     number density of $10^{11}-10^{16}~\rm{cm^{-3}}$. If the materials are inside the GRB jets
launched by the central engine (e.g., Lazzati et al. 1999),
the emission lines produced by some heavy elements,
if they are detected, can provide vital information on the GRB nucleosynthesis.

We assume that the electron density is a constant along the line of sight
in a range of $10^{11}-10^{16}~\rm{cm^{-3}}$. We take two constant numbers
 as examples and the results are shown in Figure 1.
 However, the density distribution can be complex.
 For instance, the density around the GRB jets may not only be radially dependent,
 but also angle dependent. The density distribution was discussed in some
 structured jet-cocoon models
     \citep{Ramirez+2002, Lazzati+2005, Lazzati+2017, deColle+2018, Ito+2019, Lazzati+2019}. The jet is
      collimated in a small angle by the hot cocoon pressure.
      According to some numerical simulations
       \citep{Lazzati+2017, Lazzati+2018, Lazzati+2019},
the bulk Lorentz factor in the cocoon ranges from $10-1$ in the angle of $10^\circ -40^\circ$.
 Assuming that the bulk
motion of the line-production region is the main reason for the line broadening,
we have two FWHM values
$\Delta\nu /\nu_0=0.1$ and $\Delta\nu/\nu_0=0.5$. The values are adopted in Section 3.
It implies that the emission-line region is more complicated.
The emission line is wide in the on-axis case, while the emission line is narrow in the off-axis case.
However, the emission-line photons along the line of sight can pass through the regions with different electron number densities. In general, it still seems hard to detect GRB emission lines in the X-ray band after $10^4$\,s,
and the thermal continuum from the cocoon can be observed.

A thermal component has been identified in some GRB X-ray afterglows, and
   it is usually fitted by a blackbody shape \citep{Sparre+2012,Starling+2012,Valan+2018}.
   The blackbody feature is usually shown below 1\,keV. In this paper, the blackbody shape due to the Fe K$\alpha$ line evolved
   at a very late stage and is shown in a range of $2-4$\,keV. However, the blackbody shape seems very broad, and it is extended below
   1\,keV in some cases. Besides the Fe K$\alpha$ line, some other emission lines may also suffer the down-Comptonization effect. Therefore,
   the final blackbody shape can be the result of the overlaps by several emission lines with the down-Comptonization.
    Therefore, we speculate that the thermal component identified in some GRB X-ray spectra may contain
 the contribution from the emission lines evolved at a very late stage.
 In other words, the thermal component in the GRB X-ray spectra might be the relic of the X-ray emission lines.
    In our former work \citep{Liu+2019}, the X-ray thermal component
     of the ultralong GRB 130925A observed at 1.8\,days by both
    Swift-XRT and NuSTAR can be well explained by the bremsstrahlung
    radiation in a metal-rich and dusty environment. We note that the thermal component is shown beyond 10 keV.

  \section{Conclusions}
 Aiming to interpret the low significant detection of the GRB X-ray emission line, we
 utilize the down-Comptonization mechanism and obtain the time evolution of the emission line
  by solving the Kompaneets equation. We compare the effects of the electron number density, electron temperature, and the initial FWHM line profile on the time evolution of the emission line. We find that the emission line evolves faster in cases of larger number density and higher temperature. The emission photons finally keep equilibrium with electrons and distribute as a blackbody shape.
    Compared with the observational results of the emission line, we suggest that when the number density of environment is larger than $10^{12}$\,$\rm cm^{-3}$,
    the line-like feature is hardly recognized in the GRB X-ray emission after $100$\,s.
    Our results also imply that the emission line at a very late evolutional stage
     might be identified as a thermal component in the GRB X-ray spectrum.
 \acknowledgments
We appreciate the referee's helpful suggestions. We thank S. Campana for his useful comments. J.-Y.L. appreciates discussions with Dr. Q.-S. Zhang and X.-L.Yang. J.-Y.L. acknowledges the financial support of the National Natural Science Foundation of China 11303086.
J.M. is supported by the National Natural Science Foundation of China 11673062, the Hundred Talent Program of Chinese Academy of Sciences, and the Oversea Talent Program of Yunnan Province.

\clearpage

\begin{figure}
 \includegraphics[scale=0.45]{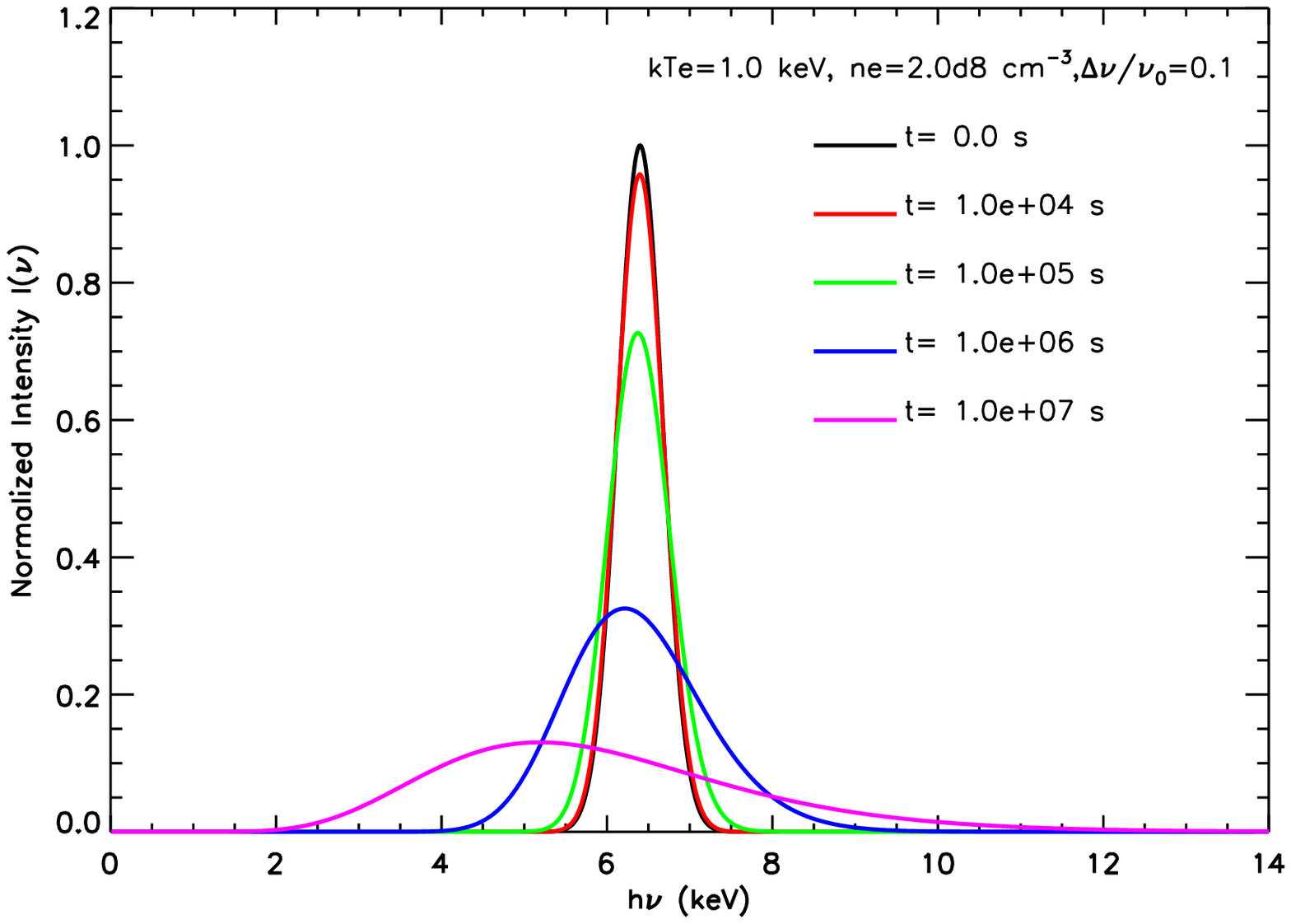}%
 \includegraphics[scale=0.45]{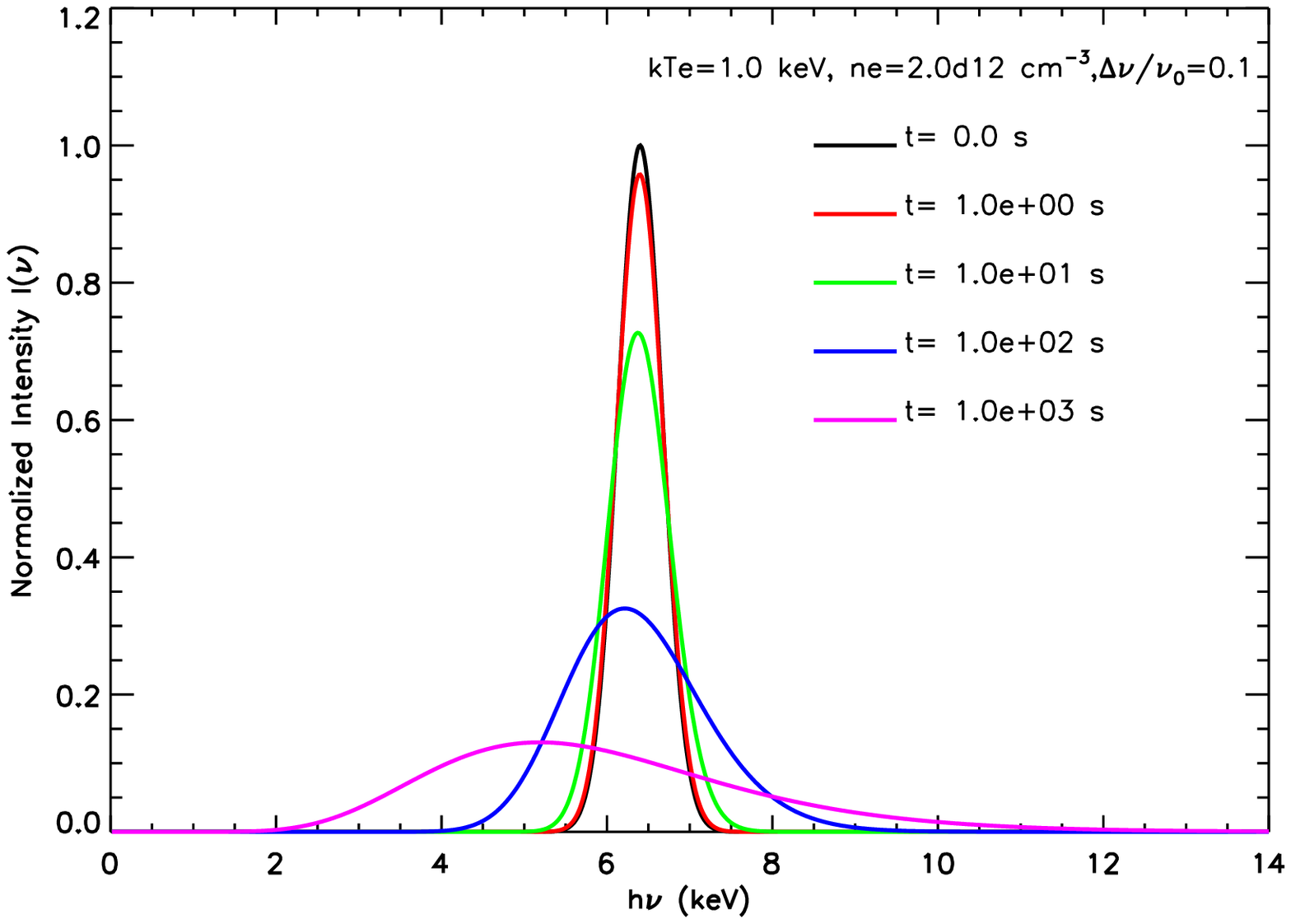}\\%
\includegraphics[scale=0.45]{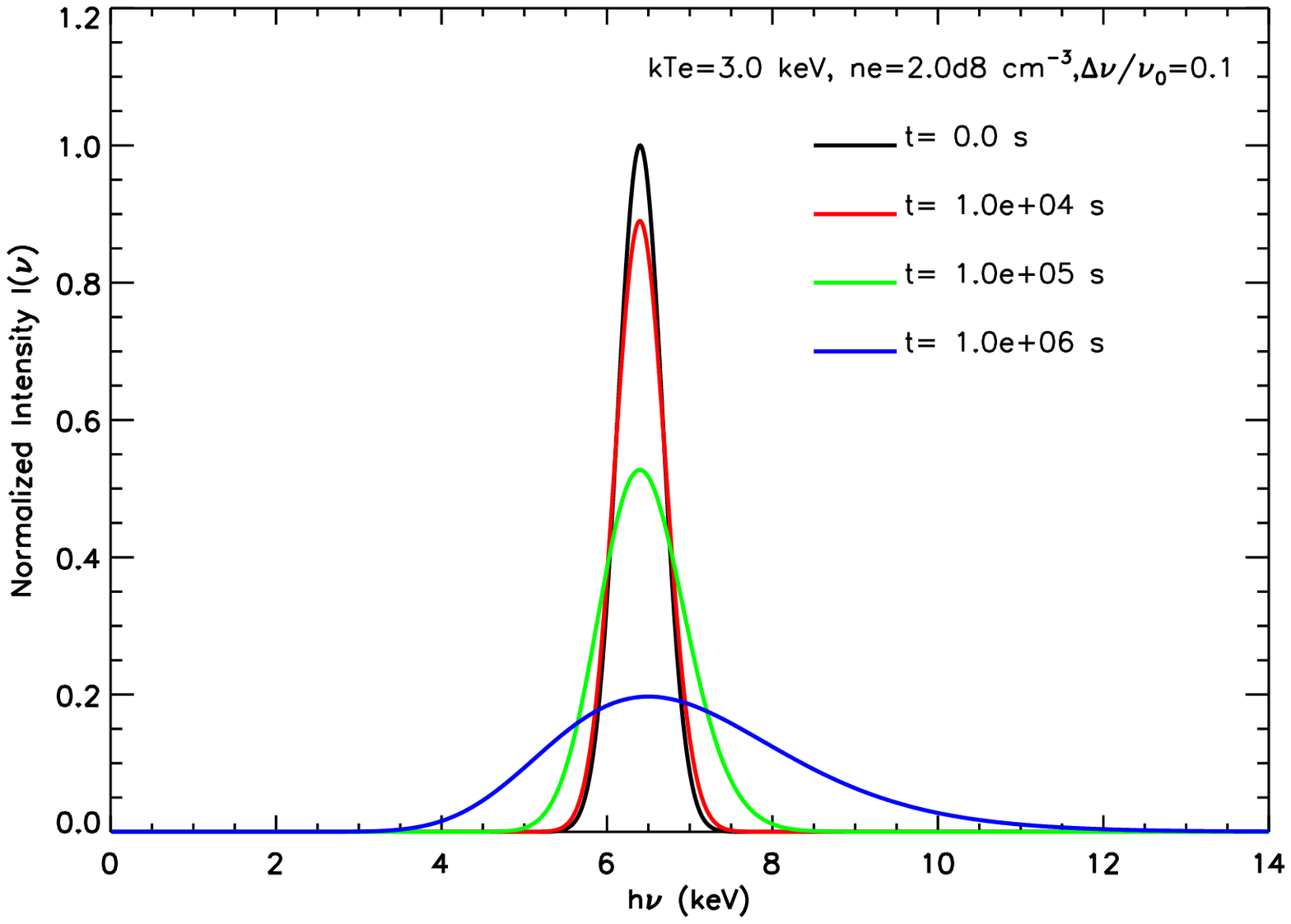}
\includegraphics[scale=0.45]{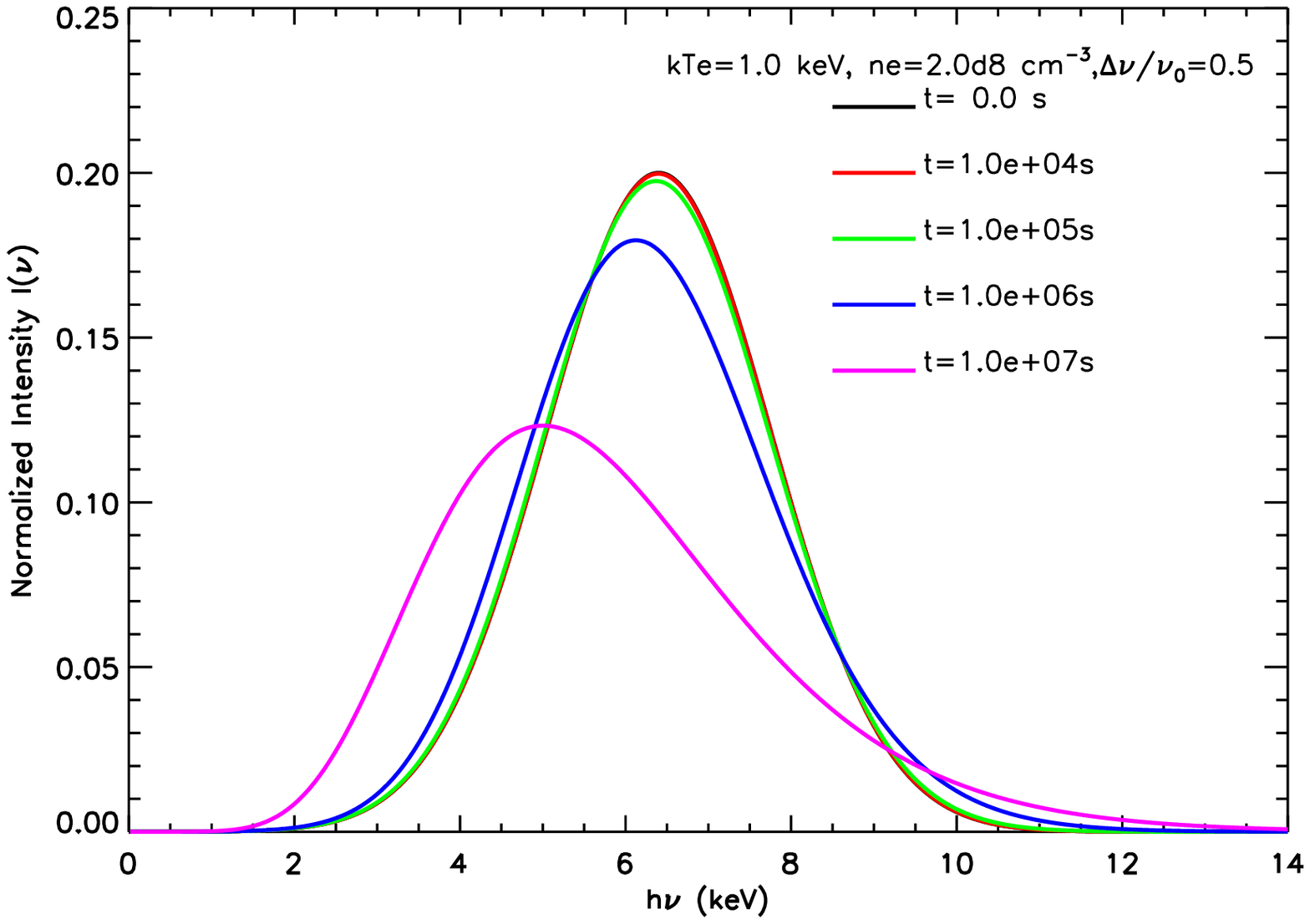}
\caption{The evolutions of the Fe K$\alpha$ line. Top-left panel: $kT_e=1.0$\,keV, $n_e=2.0\times10^8$\,$\rm cm^{-3}$, and $\Delta\nu/\nu_0 =0.1$.
Top-right panel: $kT_e=1.0$\,keV, $n_e=2.0\times10^{12}$\,$\rm cm^{-3}$, and $\Delta\nu/\nu_0 =0.1$.
Bottom-left panel: $kT_e=3.0$\,keV, $n_e=2.0\times10^8$\,$\rm cm^{-3}$, and $\Delta\nu/\nu_0 =0.1$.
Bottom-right panel: $kT_e=1.0$\,keV, $n_e=2.0\times10^8$\,$\rm cm^{-3}$, and $\Delta\nu/\nu_0 =0.5$.}\label{fig1}
\end{figure}

\clearpage

\end{document}